\documentclass[tighten,twocolappendix,twocolumn]{aastex631}

\usepackage{amsmath,footmisc}
\usepackage[T1]{fontenc}

\def\bhspin{a_*}
\def\sgra{Sgr\,A$^*$}

\defcitealias{PaperI}{EHTC~I}
\defcitealias{PaperV}{EHTC~V}
\defcitealias{PaperVI}{EHTC~VI}
\defcitealias{PaperVII}{EHTC~VII}
\defcitealias{PaperVIII}{EHTC~VIII}
\defcitealias{Paper2}{Paper II}
\defcitealias{Paper3}{Paper III}

\begin{document}

\title{Measuring Black Hole Light Echoes with Very Long Baseline Interferometry}

\shorttitle{Interferometric Black Hole Light Echoes}

\author[0000-0001-6952-2147]{George~N.~Wong}
\email{gnwong@ias.edu}
\affiliation{School of Natural Sciences, Institute for Advanced Study, 1 Einstein Drive, Princeton, NJ 08540, USA}
\affiliation{Princeton Gravity Initiative, Princeton University, Princeton NJ 08544, USA}

\author[0000-0003-2342-6728]{Lia~Medeiros}
\altaffiliation{NASA Hubble Fellowship Program, Einstein Fellow}
\affiliation{Department of Astrophysical Sciences, Princeton University, Princeton NJ 08544, USA}

\author[0000-0001-9528-1826]{Alejandro C\'{a}rdenas-Avenda\~{n}o}
\affiliation{Princeton Gravity Initiative, Princeton University, Princeton NJ 08544, USA}
\affiliation{Computational Physics and Methods (CCS-2), Los Alamos National Laboratory, Los Alamos, NM 87545, USA}
\affiliation{Center for Nonlinear Studies (CNLS), Los Alamos National Laboratory, Los Alamos, NM 87545, USA}

\author[0000-0001-5603-1832]{James~M.~Stone}
\affiliation{School of Natural Sciences, Institute for Advanced Study, 1 Einstein Drive, Princeton, NJ 08540, USA}

\begin{abstract}
Light passing near a black hole can follow multiple paths from an emission source to an observer due to strong gravitational lensing. Photons following different paths take different amounts of time to reach the observer, which produces an echo signature in the image. The characteristic echo delay is determined primarily by the mass of the black hole, but it is also influenced by the black hole spin and inclination to the observer. In the Kerr geometry, echo images are demagnified, rotated, and sheared copies of the direct image and lie within a restricted region of the image. Echo images have exponentially suppressed flux, and temporal correlations within the flow make it challenging to directly detect light echoes from the total light curve. In this paper, we propose a novel method to search for light echoes by correlating the total light curve with the interferometric signal at high spatial frequencies, which is a proxy for indirect emission. We explore the viability of our method using numerical general relativistic magnetohydrodynamic simulations of a near-face-on accretion system scaled to M87-like parameters. We demonstrate that our method can be used to directly infer the echo delay period in simulated data. An echo detection would be clear evidence that we have captured photons that have circled the black hole, and a high-fidelity echo measurement would provide an independent measure of fundamental black hole parameters. Our results suggest that detecting echoes may be achievable through interferometric observations with a modest space-based very long baseline interferometry mission.
\end{abstract}


\section{Introduction}

Recent very long baseline interferometry (VLBI) observations by the Event Horizon Telescope (EHT) collaboration have revealed that the supermassive black holes in the centers of our galaxy, Sagittarius A$^*$ (\sgra), and the nearby elliptical galaxy M87 produce bright rings of emission at $230\,{\rm GHz}$. EHT analyses have informed constraints on the black holes' masses and angular momenta, properties of the surrounding accretion flows like magnetic field structure and accretion rate, and deviations from the Kerr metric \citep{eht_m87_1,eht_m87_4,eht_m87_5,eht_m87_6,eht_m87_7,eht_m87_8,eht_m87_9,eht_sgra_1,eht_sgra_4,eht_sgra_5,eht_sgra_6,eht_sgra_7,eht_sgra_8}. The EHT's demonstrated technological capability to recover resolved, polarimetric images of black holes at event horizon scales has spurred efforts to increase the size of the telescope array and develop simultaneous multi-frequency receivers. These advances promise improved image fidelity and science from multi-wavelength data. There has also been growing interest in designing new space-based interferometers, which would enable probes of the complicated emission ring substructure and improve prospects for black hole spin measurements and further tests of general relativity \citep{kudriashov_2021_ehi,gurvits_2021_theza,kurczynski_2022_ehe,johnson_2024_bhex}.

The structure of the observed emission ring is influenced by both the complex astrophysics of the accretion flow and the geometry of the black hole spacetime. Many of the contemporary analysis methods used on EHT data rely on image reconstruction, fitting geometric models to the interferometric products, and comparing simulations to the data \citep{eht_m87_4,eht_m87_5,eht_m87_6,eht_m87_8,eht_sgra_4,eht_sgra_5,eht_sgra_6,eht_m87_9,eht_sgra_8}. However, there is also significant interest in 
finding probes of the spacetime geometry that are less influenced by astrophysical uncertainties. Strong gravitational lensing near the event horizon allows photons to travel around the black hole multiple times. On the image plane, the boundary of the black hole shadow (or critical curve) is defined by the critical impact parameter separating photons on geodesics that come from infinity from those on geodesics that pass through the event horizon. The size and shape of the black hole shadow are independent of the astrophysical emission model and therefore act as a direct probe of the spacetime geometry. 

Although the shadow boundary cannot be observed directly, it is often traced by a bright ring of emission; the relationship between the sizes of the observed ring and the black hole shadow has been calibrated with semianalytic models and simulations \cite{eht_sgra_6}. There is significant interest in using the size and shape of the ring to infer bounds on spin, and some proposals to test the Kerr hypothesis also rely on looking for deviations in the ring shape
\citep{falcke_2000_shadow,takahashi_2004_spinshape,bambi_2009_shadow,hioki_2009_spinshadow,amarilla_2010_chernsimonsshadow,tsukamoto_2014_shadow,amarilla_2013_kaluzaklein,younsi_2016_shadowcalc,mizuno_2018_shadowtestgr,medeiros_2020_nonkerrshadow,olivares_2020_bosonstar,johnson_2020_universal,wielgus_2020_wormholes,psaltis_2020_ehtgrshadow,kocherlakota_2021_ehtcharges,eht_sgra_6,staelens_2023_phringsbeyondgr,lupsasca_2024_guidegr,cardenasavendano_2024_lensingbands}.

The above methods focus primarily on image features, but the spacetime geometry also produces signatures in the time domain. Since an emission source can radiate in multiple directions simultaneously, different photons emitted by the same source but in different directions can orbit the black hole different numbers of times before reaching an observer. This sequence of time-delayed photons produces a characteristic light echo whose delay period is determined primarily by the black hole mass: the difference in arrival times between direct ($n=0$) and singly lensed indirect ($n=1$) emission is of order the time it takes for light to travel around the far side of the black hole.

Since the image contains contributions from direct and all orders of indirect photons, searching for echo features in the (auto)correlation function of the light curve or resolved movies could in principle be used to constrain properties of the spacetime (e.g., \citealt{broderick_2005_hotspots,fukumura_2008_lightechoes,moriyama_2015_echoes,moriyama_2019_echoessynthetic,wong_2021_glimmer,hadar_2021_autocorrelations,cardoso_2021_matterappearance,andrianov_2022_flares,hadar_2023_spectrotemporal}). However, \citet{cardenasavendano_2024_noechoes} showed that direct detection of the echo signature from the total image light curve is challenging, since the flux in the $n=1$ image is exponentially suppressed relative to the flux in the $n=0$ image, and correlations within the underlying accretion flow can make it difficult to disentangle the echo signature.

A more productive detection strategy might try to isolate the $n=0$ and $n=1$ light curves. In this paper, we propose to use interferometric time series data at short and long baselines to obtain proxies for the $n=0$ and $n=1$ light curves independently. Since the $n=1$ image is characteristically thinner than the $n=0$ image and dominates the signal at high spatial frequencies, variations in the visibility amplitudes at long baselines can be used to infer information about variations in the indirect light curve.

We test the viability of the proposed echo detection method using numerical general relativistic magnetohydrodynamics (GRMHD) simulations of accretion flows. We specialize our study to the M87 black hole, where the orientation of a large scale radio jet has been used to infer that the central black hole spins with an inclination of $17^{\circ}$ away from the observer's line of sight \citep{walker_2018_m87jet,eht_m87_5}. Our focus on a nearly face-on observer allows us to restrict our attention to the simplest case (we explore the echo signature at higher inclinations in Appendix~\ref{app:inc_asymp}). By computing the correlation function between the total light curve and time series visibility amplitudes at longer baselines, we find an echo signature that is consistent with the time delay predicted by general relativity.

The remainder of this paper is organized as follows. In Section~\ref{sec:theory}, we describe the theory of black hole light echoes. In Section~\ref{sec:observations}, we use a simulated black hole movie to demonstrate the viability of performing an interferometric echo measurement and discuss several observational considerations. We conclude with a brief discussion and summary in Section~\ref{sec:discussion}. We discuss technical definitions and implementation details in the Appendices.

\section{Black Hole Light Echoes}
\label{sec:theory}

In this section, we explore the theory of black hole light echoes. We begin with a review of the Kerr spacetime including its null geodesics and the $n$-ring subimage decomposition. We continue with a study of echoes in a numerical simulation and describe a simple semi-analytic echo model. We use the semi-analytic model to show how different astrophysical fluid prescriptions lead to the presence (or absence) of echoes that can be detected directly from the light curve.

\subsection{The Kerr geometry}

Isolated black holes in general relativity are described by the Kerr metric, which depends only on mass and angular momentum.\footnote{Black holes may also be charged, but it is unlikely that astrophysically relevant objects have dynamically important charge.} In what follows, we overview the Kerr geometry and the equations that govern the null geodesic trajectories. {We then list several quantities associated with null geodesics that are useful in describing the echo signature.

In this section, we use geometrized units with $G = c = 1$ and describe the black hole angular momentum $J$ in terms of the dimensionless spin parameter $\bhspin \equiv J/M^2$, where $M$ is the mass of the black hole. We also set $M=1$ for clarity, although we restore constants and the mass dependence in the rest of the paper. In Boyer-Lindquist coordinates, $x^\mu = \left( t, r, \theta, \phi \right)$, the Kerr line element is \citep{bardeen_1972_kerr}
\begin{align}
ds^2 =& - \left(1 - \dfrac{2r}{\Sigma} \right) dt^2 - \dfrac{4 \bhspin r \sin^2 \theta}{\Sigma} dt\, d\phi + \dfrac{\Sigma}{\Delta} dr^2 \nonumber \\
& + \Sigma d\theta^2 + \dfrac{\left(r^2 + \bhspin^2\right)^2 - \Delta \bhspin^2 \sin^2 \theta}{\Sigma} \sin^2 \theta \, d\phi^2,
\end{align}
with
\begin{align}
\Sigma(r) &\equiv r^2 + \bhspin^2 \cos^2 \theta, \\ 
\Delta(r,\theta) &\equiv r^2 - 2 r + \bhspin^2.
\end{align}

The trajectory of a particle with four-momentum $p^\mu$ is given by the solution to the geodesic equations
\begin{align}
\dfrac{d x^\mu}{ds} &= p^\mu \\
\dfrac{d p^\mu}{ds} &= - {\Gamma^\mu}_{\alpha\beta} p^\alpha p^\beta,
\end{align}
where ${\Gamma^\mu}_{\alpha\beta}$ is a Christoffel symbol and $s$ is an affine parameter. Photons follow null geodesics through spacetime, and the photon four-momentum is often written as the null four-wavevector $k^\mu$.

\vspace{0.5em}

The Kerr spacetime admits four conserved quantities: an invariant mass $\mu$, energy at infinity $E = -p_t$, azimuthal angular momentum $L_z = p_\phi$, and the Carter constant  (a generalized total angular momentum), defined by
\begin{align}
\mathcal{Q} = p_\theta^2 + \cos^2 \theta \left( \bhspin^2 \left(\mu^2 - p_t^2\right) + p_\phi^2 /\sin^2 \theta\right).
\end{align}
Null geodesics have $\mu = 0$. They are also independent of the magnitude of $E$ and can therefore be fully characterized by their energy-rescaled angular momentum $\Phi = L_z / E$ and normalized Carter constant $Q = \mathcal{Q}/E^2$. The geodesic equations admit closed-form explicit solutions in terms of integrals of motion as well as explicit parameterized trajectories \citep[e.g.,][]{li_2005_thindiskspectrum,dexter_2009_geokerr,gralla_2020_kerrlensing}, but they are often solved numerically \citep{gold_2020_grrtcomp,prather_2023_polcomp}.

\vspace{0.5em}

Solving for null geodesics exposes the existence of spherical (constant $r$) bound orbits close to the black hole. The set of spherical orbits comprises the \emph{photon shell}. For a particular black hole spin, the spherical photon orbits can be uniquely parameterized by their radii, which lie continuously in the range $r_- \le r \le r_+$, with
\begin{align}
r_{\rm ph, \pm} &= 2 \left( 1 + \cos \left(\dfrac{2}{3} \arccos \left( \pm \bhspin \right) \right) \right).
\end{align}
The extremal orbits at $r_-$ (prograde) and $r_+$ (retrograde) lie in the midplane at $\theta = \pi/2$, but others oscillate between different latitudes. 
The constants of motion for the spherical orbits can be expressed in terms of $r$ as
\begin{align}
\Phi &= -\dfrac{r^3 - 3r^2 + \bhspin^2 r + \bhspin^2}{\bhspin \left(r-1\right)}, \\
Q &= - \dfrac{r^3 \left( r^3 - 6 r^2 + 9r - 4 \bhspin^2\right)}{\bhspin^2 \left(r-1\right)^2}.
\end{align}
Between $r_{-}$ and $r_+$ lies the polar orbit
\begin{align}
r_{\rm polar} = 1 + 2 \sqrt{1 - \dfrac{1}{3}\bhspin^2} \cos\left( \dfrac{1}{3} \arccos \dfrac{1 - \bhspin^2}{\left(1 - \frac{1}{3}\bhspin^2\right)^{3/2}} \right),
\end{align}
which has $\Phi = 0$.
A detailed discussion of photon orbit oscillations and their image-domain observational consequences is provided in \citet{wong_2021_glimmer}.

\vspace{0.5em}

Spherical photon orbits are unstable in Kerr, and photons following geodesics whose paths are slightly misaligned will either fall through the event horizon or escape to infinity.
This exponential instability is described by a Lyapunov exponent $\gamma$, which quantifies how the deviation $\delta r$ between a geodesic's radial position and the position of the true spherical orbit grows (or decreases) after $n$ latitudinal half-cycles (see, e.g., \citealt{yang_2012_qnmkerr,johnson_2020_universal})
\begin{align}
\delta r_n = e^{\pm \gamma n} \, \delta r_0.
\end{align}
Here we take a latitudinal half-cycle to be the segment of motion from $\theta=\pi/2$ to the extremum in $\theta$ back to $\theta=\pi/2$.
The Lyapunov exponent is
\begin{align}
\gamma = \dfrac{4}{\bhspin \sqrt{-u_-^2}} \sqrt{r^2 - \dfrac{r\,\Delta}{\left(r-1\right)^2}}\ K\left(\dfrac{u_+^2}{u_-^2}\right),
\end{align}
where $K$ is the complete elliptic integral of the first kind\footnote{We use the square of the elliptic modulus as the parameter for the elliptic integral.} and where the roots of the latitudinal potential are
\begin{align}
u_\pm^2 \equiv \dfrac{\bhspin^2 - Q - \Phi^2 \pm \sqrt{\left(Q + \Phi^2 - \bhspin^2\right)^2 + 4 \bhspin^2 Q}}{2 \bhspin^2}.
\end{align}
The time delay associated with one latitudinal half-cycle can also be expressed in terms of elliptic integrals
\begin{align}
\label{eq:time_delay}
\Delta t = \  & 2 \, \dfrac{\left(r^2 + \bhspin^2\right)^2 - 2 \bhspin \Phi r - \bhspin^2 \Delta}{\bhspin \Delta \sqrt{-u_-^2}} K \left( \dfrac{u_+^2}{u_-^2} \right) \nonumber \\
& - 2 \bhspin \sqrt{-u_-^2 u_+^2} \left[ K \left(\dfrac{u_+^2}{u_-^2} \right) - E \left( \dfrac{u_+^2}{u_-^2} \right) \right],
\end{align}
where $E$ is the complete elliptic integral of the second kind.

\subsection{Subrings and the geometry of echoes}
\label{sec:echo_theory}

\begin{figure*}[th!]
\centering
\includegraphics[trim={0 2cm 0 0},clip,height=23em]{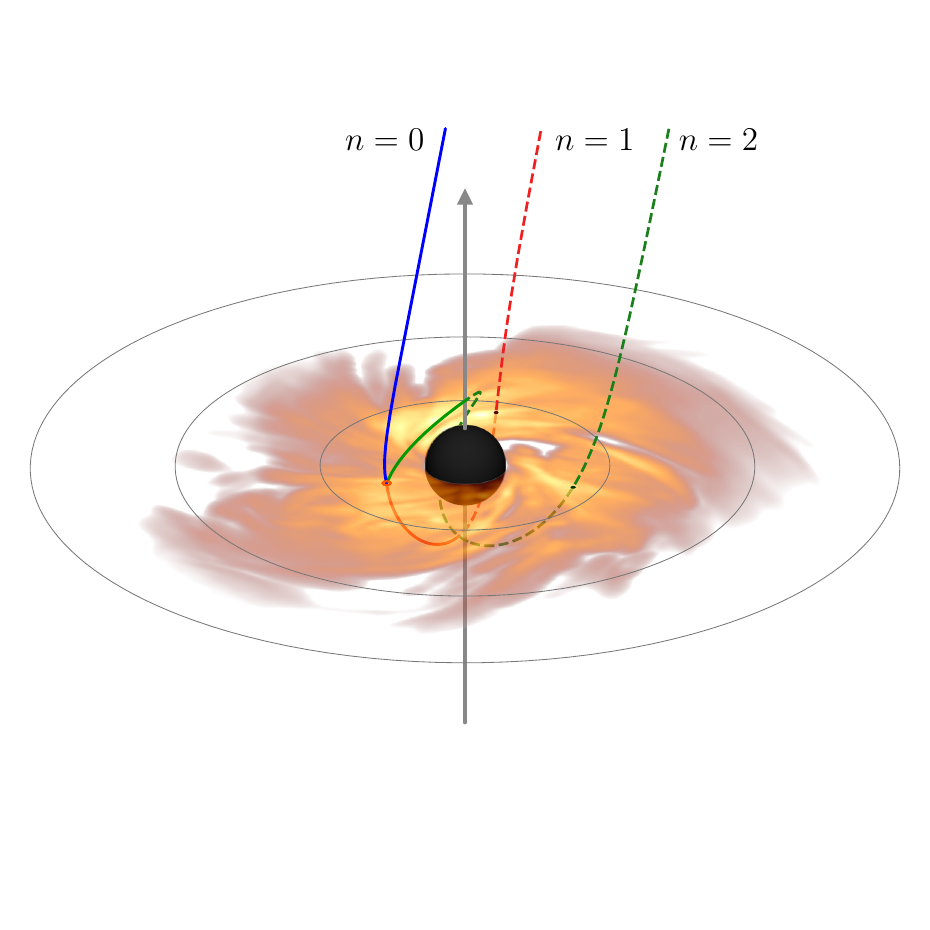} \qquad
\includegraphics[height=22em]{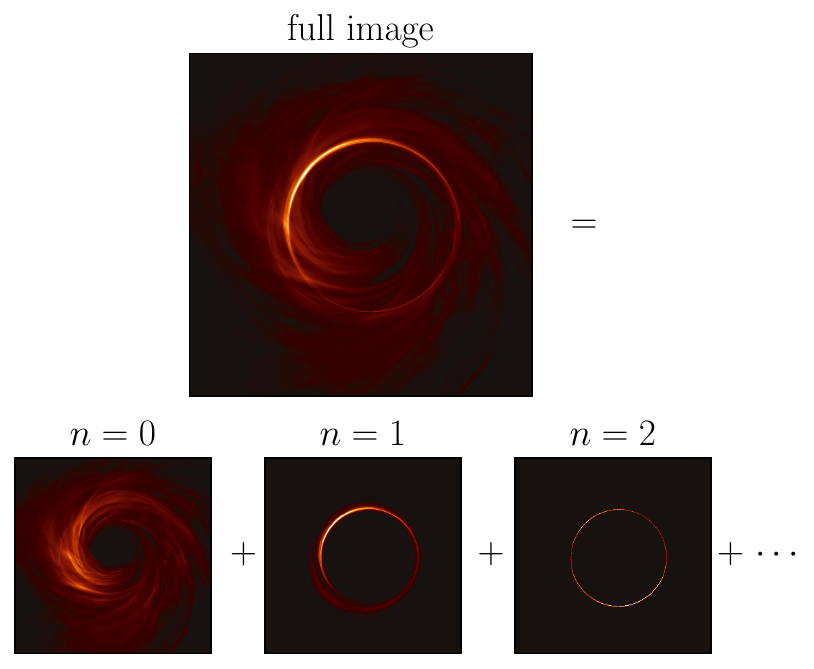}
\caption{
\textit{(left)} Schematic illustration of a black hole accretion disk illustrating how photons can follow different paths between a single point in the flow and the observer. The colored disk shows emission on an equatorial slice in a snapshot from a GRMHD simulation, with a linear colormap showing the number of photons that were produced from that part of the midplane and then captured by the observer. The black hole has $\bhspin=0.9375$ with spin axis pointing up and its event horizon is depicted as a filled black sphere. The observer is located at an inclination of $i=17^{\circ}$ to the right of the spin axis. We highlight a single emission point left of the black hole with a red circle. The path followed by the direct emission ($n=0$) is shown in blue, by the first order indirect emission ($n=1$) is shown in red, and by the second order ($n=2$) in green. Emission originating from any point in the solid colored geodesic path would appear in the corresponding subimage (i.e., emission along the solid red segment would also appear in the $n=1$ subimage).
\textit{(right)} Ray-traced image of the GRMHD snapshot on the left: the total observed flux (top) can be decomposed into the $n=0, 1, 2, \ldots$ subimages (bottom). The $n=0$ sub-image does not contain the sharp bright ring feature seen in the total image, while the higher-order subimages are dominated by that feature, which becomes thinner and sharper as $n$ increases.
}
\label{fig:nring_splash_cartoon}
\end{figure*}

The photons that reach an observer arrive at the image plane with a set of screen coordinate impact parameters $(\alpha, \beta)$. Following \citet{bardeen_1973_kerrgeo}, we align the $\alpha=0$ axis with the projection of the black hole spin axis. In this paper, we use dimensionless polar image coordinates 
\begin{align}
& \rho = \sqrt{\alpha^2 + \beta^2}, \\
& \tan \varphi = \dfrac{\beta}{\alpha},
\end{align}
rather than the Cartesian coordinates of Bardeen. For a distant observer at radius $r_{\rm obs}$ and inclination $i$, the conserved quantities of a photon are related to the impact parameters as
\begin{align}
& \rho = \dfrac{1}{r_{\rm obs}} \sqrt{\bhspin^2 \left( \cos^2 i - u_+^2 u_-^2 \right) + \lambda^2}, \\
& \cos \varphi = - \dfrac{\lambda}{r_{\rm obs} \, \rho \, \sin i} .
\end{align}

As stated above, photons emitted at a single spacetime event $x^\mu$ near a black hole may take multiple paths to an observer. Distinct geodesics have different conserved quantities and will in general arrive at different locations on the observer's screen with different time delays (as illustrated in Figure~\ref{fig:nring_splash_cartoon}). The multiple paths give rise to an infinite sequence of lensed \emph{subimages} of the universe \citep{johnson_2020_universal}. Because the conditions for a geodesic to wrap $n$ times around the black hole become increasingly restrictive with increased $n$, each higher-order subimage lies within an increasingly thinner region on the observer's plane (the $n$th \emph{lensing band}, \citealt{gralla_2020_kerrlensing}; right panel of Figure~\ref{fig:nring_splash_cartoon}). In the limit of $n \to \infty$, the sequence of subimages converges to the critical curve, i.e., the geometric boundary of the black hole shadow.

\begin{figure*}[th!]
\centering
\includegraphics[width=\textwidth]{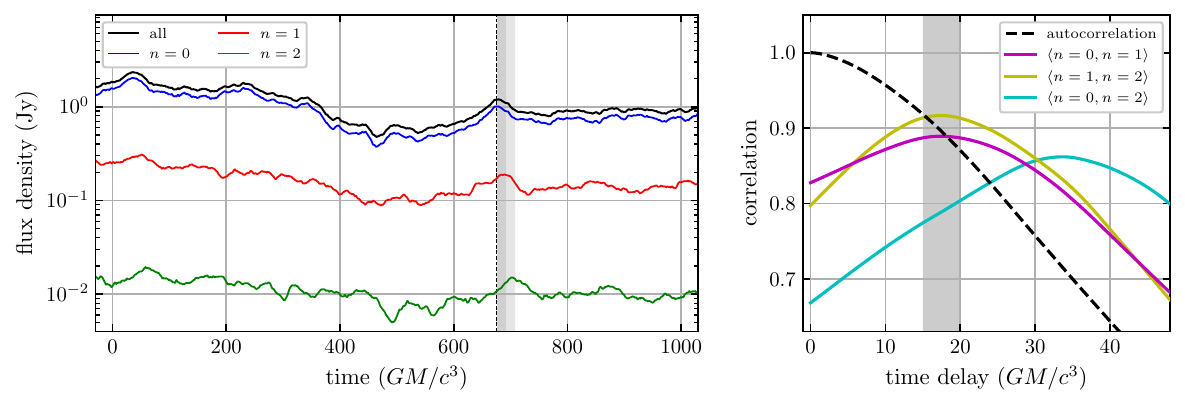}
\caption{\textit{(left}) Light curves from the same $\bhspin = 0.9375$ MAD GRMHD simulation shown in Figure~\ref{fig:nring_splash_cartoon}. The total light curve (black) has been decomposed into light curves for the $n=0$ direct emission (blue), $n=1$ indirect emission (red), and $n=2$ doubly indirect (green) subimages. The flux decreases by about an order of magnitude between different values of $n$. The three light curves are similar, but they are offset by about $16\,GM/c^3$, which is the approximate amount of time it takes for light emitted near the black hole to travel around its far side. As an example, the peak in the direct emission at the vertical black dashed line is delayed in the first order indirect emission ($n=1$; the dark gray shaded region denotes a delay of $16\,GM/c^3$) and further delayed in the second order light curve ($n=2$; the light gray shaded region denotes a delay of $32\,GM/c^3$).
\textit{(right)} Correlations between different light curves as a function of time delay. The dashed black curve shows the autocorrelation of the total light curve, and the colored curves correspond to the correlation between the $n=0$ and $n=1$ emission (magenta), $n=1$ and $n=2$ (yellow), and $n=0$ and $n=2$ (cyan). Although the autocorrelation of the total light curve does not show an echo peak, echo peaks can be clearly seen in the correlations between the subimage light curves. The peak time delay between emission separated by one order is between $15$ and $20\,GM/c^3$ (the dark gray shaded interval), while the peak time delay between the $n=0$ and $n=2$ light curves is about twice as long, as expected.
}
\label{fig:lightcurve_decomposition}
\end{figure*}

The geodesic segments connecting a source location to an observer can be ordered according to the time it would take for a photon to travel along them.\footnote{Sources that lie within the caustic surface for some observer will have multiple geodesics with the same travel time delay to the observer (see \citealt{rauch_1994_caustics} and \citealt{bozza_2008_kerrcaustics} for discussions of caustics in Kerr). We neglect the presence of caustics in this schematic description of echoes; however, we self-consistently account for their presence in all calculations that appear in this paper.} This labeling yields the time-ordering definition for subimages that we use in this paper: photons that follow the shortest path to the observer belong to the $n=0$ subimage, photons that follow the second shortest path belong to the $n=1$ subimage, and so on. This definition has several appealing properties including that each emission source appears once per subimage and a given point in space $x^i$ and subring index $n$ uniquely identifies a geodesic connecting the point and the observer (again neglecting the influence of caustics). See Appendix~\ref{app:subimages} for a discussion of how this time-ordering definition of subimages differs in detail compared to the turning point or midplane-crossing ones often used in other work (e.g., \citealt{johnson_2020_universal}; see also \citealt{zhou_2024_forwardkerr} for some nuances associated with image degeneracy in turning point definitions).

The $n$th subimage of a source is typically demagnified, delayed, and rotated with respect to the $(n-1)$th subimage, although the details are less clean for small $n$ and for sources off the midplane. In this paper, we are most interested in the time delay between subsequent subimages, which results in the light echo. For a face-on observer, the characteristic echo delay is well approximated by the time it takes for light to complete one half-orbit around the black hole within the photon shell. The precise analytic form of this time delay as a function of spin comes from Equation~\ref{eq:time_delay} and typically takes on values around $\sim 16\,GM/c^3$. As we will see in the following section, the observed echo delay(s) depend on the emission source and can be more complicated for observers at higher inclinations; however, the average light travel time is still well described by this approximation.

\subsection{Echoes in a numerical accretion flow simulation}

To test whether the echo signature produced from a turbulent accretion flow might be detectable, we generate high cadence, high resolution black hole movies by raytracing GRMHD simulations. We specialize our analysis to the M87 black hole, which is one of the primary EHT sources. M87 is a particularly appealing target because it is large on the sky, varies on days-long timescales, and is inferred to be viewed nearly face on \citep{eht_m87_5,eht_m87_8}. In Sections~\ref{sec:theory}~and~\ref{sec:observations}, we focus on one fluid simulation produced with the {\tt athenak} code (J.~Stone et al.~in preparation), but we investigate the robustness of our results in Section~\ref{sec:discussion} with separate simulations produced with the alternative GRMHD code KHARMA \citep{prather_2021_iharm3d}.

\begin{figure}[th!]
\centering
\includegraphics[width=\linewidth,clip,trim={0 0 -2em 0}]{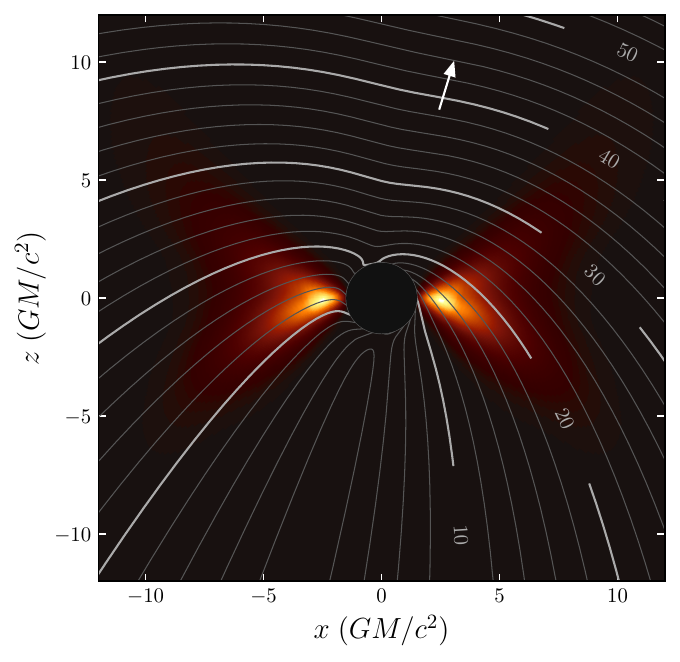}
\caption{
Location of observed emission relative to the time delay between $n=0$ and $n=1$ geodesics connecting that location to the observer. The data used for this figure come from the same $\bhspin = 0.9375$ simulation used in Figures~\ref{fig:nring_splash_cartoon}~and~\ref{fig:lightcurve_decomposition}, which observe the accretion flow at an inclination of $i=17^\circ$ (indicated by the white arrow). The color map corresponds to the azimuthally averaged emissivity plotted in the $y=0$ plane and is shown in a linear scale. The contours depict isodeltachrones, surfaces where the $n=1$ emission would arrive at the observer with a fixed time delay relative to the $n=0$ emission. Most emission comes from regions where the time delay is in the range $10-20\,GM/c^3$. 
}
\label{fig:isodeltachrones_main}
\end{figure}

The EHT analyses infer that the M87 accretion system is in the magnetically arrested disk (MAD; \citealt{bisnovatyi_1974_madstar,igumenshchev_2003_mad,narayan_2003_mad}) state and contains a spinning black hole \citep{eht_m87_5,eht_m87_8,eht_m87_9}. We target our GRMHD simulations to such a MAD system and set the spin of the central black hole to $\bhspin = 0.9375$. We initialize from the equilibrium torus solution of \citet{fishbone_1976_torus} with an inner disk radius of $r_{\rm in} = 20\,GM/c^2$ and a radius of pressure maximum $r_{\rm max} = 41\,GM/c^2$. We set the fluid adiabatic index to $\hat{\gamma} = 13/9$. These values are consistent with the EHT analysis \citep{eht_m87_5,eht_m87_8,eht_m87_9}. 

Our simulation is performed on a Cartesian mesh extending to $\pm 1024\,GM/c^2$ with eight levels of mesh refinement, where the highest resolution has $16$ grid cells per $GM/c^2$ and extends to $\pm 8\,GM/c^2$ in each of the three directions. We evolve the flow for a total duration of $20,000\,GM/c^3$, although we discard the initial $10,000\,GM/c^3$ of evolution to avoid the influence of the transient from the initial condition. We save the full fluid state every $\Delta t = 0.5\,GM/c^3$. A brief discussion of choices about the simulation temporal cadence is presented in Appendix~\ref{app:grmhd_cadence}.

We post-process our fluid simulations using the ``slow-light'' mode of the {\tt{ipole}} code \citep{moscibrodzka_2018_ipole,wong_2022_patoka}, which accounts for the finite speed of light by linearly interpolating between fluid snapshot files as photons travel through the computational domain. The electron temperature is determined from the fluid snapshots by the model described in \citet{moscibrodzka_2016_rhigh} with $r_{\rm low} = 1$ and $r_{\rm high} = 40$. The emission and absorption coefficients are computed following the prescription in \citet{leung_2011_rtcoeffs}. We set the observer inclination to $17^\circ$ off the black hole spin axis and rotate the position angle of the image to reproduce the observed brightness asymmetry orientation and be consistent with observations of the large-scale radio jet \citep{walker_2018_m87jet}. We use a black hole mass of $M_{\rm bh} = 6.5 \times 10^9\,M_\odot$ and set the distance between the black hole and observer to $d_{\rm src} = 16.8\,$Mpc \citep{eht_m87_5,eht_m87_8,eht_m87_9}. The mass accretion rate was chosen for each simulation to reproduce the observed target $230\,$GHz flux density of $\approx 0.65\,$Jy \citep{eht_m87_4}. More detail about the simulation pipeline can be found in \citet{wong_2022_patoka}.

By only allowing emission along certain geodesic segments, it is possible to use {\tt{ipole}} to directly simulate the appearance of different subimages. Figure~\ref{fig:lightcurve_decomposition} shows the results of performing such a subimage decomposition on the light curve from the {\tt{athenak}} simulation shown in Figure~\ref{fig:nring_splash_cartoon}. The left panel of Figure~\ref{fig:lightcurve_decomposition} shows light curves for all photons as well as for the $n=0,1,2$ emission, which each contain exponentially less flux. Notice that the higher order light curves are delayed relative to each other by $\approx 16\,GM/c^3$, as illustrated by the gray bands with the same width. The different order light curves are also not identical: Doppler effects, optical depth, and the differences in the magnetic pitch angle between different subimage orders all lead to differences in the light curves. The right panel of the figure shows the correlation functions between different pairs of light curves. We use the (Pearson) correlation coefficient for this computation 
\begin{align}
r_{XY} = \dfrac{\sum\limits_i \left(x_i - \overline{x}\right) \left(y_i - \overline{y}\right)}{\sqrt{\sum\limits_i \left(x_i - \overline{x}\right)^2}\sqrt{\sum\limits_i \left(y_i - \overline{y}\right)^2}},
\end{align}
where $X = \{x_i\}$ and $Y = \{y_i\}$ are the different light curve time series with averages $\overline{x}$ and $\overline{y}$, respectively.

The right panel of Figure~\ref{fig:lightcurve_decomposition} shows four correlations, including the total light curve with itself (i.e., the autocorrelation) as well as each of the pairs of subimage components. Evidently the echo signature is complicated. Its width and location are determined by several factors, including
\begin{itemize}
\item astrophysical correlations in the direct $n=0$ emission arising from the plasma physics and fluid dynamics,
\item the spatial distribution of emissivity, which may differ for different subimages due to the varying angles between the photon wavevector and the fluid velocity and magnetic field,
\item the position-dependent time delay between the arrivals of different subimage order photons at the observer, and 
\item the Jacobian demagnification factor between the source and the image plane, which differs for different subimage orders.
\end{itemize}

Figure~\ref{fig:isodeltachrones_main} shows the spatial distribution of emission seen by the observer in Figures~\ref{fig:nring_splash_cartoon}~and~\ref{fig:lightcurve_decomposition}. Overplotted are the \emph{isodeltachrones} for $n=0\to1$ evaluated in the $y=0$ plane. The isodeltachrones denote the surfaces along which the time delay between the $n=0$ and $n=1$ images of a source are constant, i.e., if emission were confined to a single isodeltachrone, there would be a single, sharp echo at that time delay. Although the emission peaks near the $\Delta t = 14-16\,GM/c^3$ isodeltachrones, there is still non-trivial emission over a wide range of time delays. The spin dependence of isodeltachrones is discussed in Section~\ref{sec:discussion}. Appendix~\ref{app:inc_asymp} shows several examples of isodeltachrones and emission maps for different inclinations and subimage orders.

\subsection{A simple echo model}
\label{sec:echo_model}

Figure~\ref{fig:lightcurve_decomposition} demonstrates that the autocorrelation of the total light curve does not present an obvious echo peak near the expected time delay even though analyzing the subimage components independently shows clear evidence of correlations. The absence of an echo peak in the autocorrelation can be understood via a simple model like the one introduced in \citet{cardenasavendano_2024_noechoes}.

Suppose that the $n=1$ emission is an exact, delayed, and demagnified copy of the direct $n=0$ emission and that we can neglect emission from $n>1$.
If the direct emission light curve is given by $L_0(t)$, then the total light curve will be
\begin{align}
L_{\rm tot}(t) &= L_0(t) + L_1(t) + \cdots 
\end{align}
where the light curve due to $n=1$ emission is
\begin{align}
L_1(t) &= \int\limits_0^\infty d\tau' \, \rho(\tau') \, L_0(t-\tau'),
\end{align}
and where $\rho (\tau')$ is a demagnification factor.
If we further assume that the $n=1$ emission is delayed exactly by a time $\tau$ (i.e., that all emission lies along the $\tau$ isodeltachrone), then 
\begin{align}
L_1(t) &= \int\limits_0^\infty d\tau' \, e^{-\gamma} \delta(\tau' - \tau) \, L_0(t-\tau') \\
&= e^{-\gamma} L_0(t - \tau),
\end{align}
where $\gamma$ is a characteristic Lyapunov exponent for the spacetime along the $\Delta t = \tau$ isodeltachrone.

\vspace{0.5em}

\begin{figure}[th!]
\centering
\includegraphics[width=\linewidth]{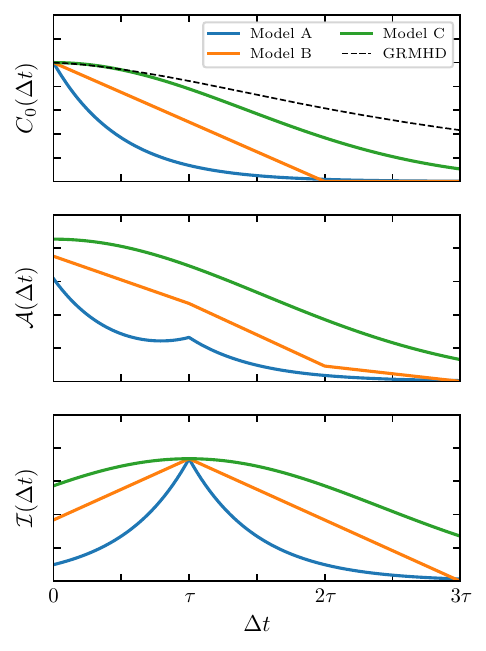}
\caption{Comparison of echo detectability for three different models of the astrophysical correlations in the direct image. We adopt the echo model of Section~\ref{sec:echo_model} and take the time delay between the $n=0$ and $n=1$ emission to be $\Delta t = \tau$ and the Lyapunov exponent to be $\gamma = \pi$. \textit{(top)} Definition of the (hypothetical) direct-emission astrophysical autocorrelation $C_0(\Delta{}t)$. For comparison, we also show the same quantity for a GRMHD simulation as a dashed black curve. \textit{(center)} The autocorrelation of the total light curves for each of the models.
When the astrophysical autocorrelation falls off more slowly, the echo signature is suppressed. \textit{(bottom)} The correlation we would measure if we could directly access the $L_0$ and $L_1$ light curves, which is a translated copy of the astrophysical correlation. For all three models, a clear echo peak is present when we correlate $L_0$ and $L_1$.
}
\label{fig:semianaytic_correlations}
\end{figure}

In order to study the effect of the astrophysical correlations, it is useful to define the autocorrelation of the direct lightcurve with itself,
\begin{align}
C_0(\Delta t) \equiv \left\langle L_0 (t)  L_0 (t + \Delta t) \right\rangle,
\end{align}
which is devoid of general relativistic effects. The autocorrelation $C_0(\Delta t)$ is an even function and its timescale is typically determined by how long it takes for fluid features to change appreciably.

\begin{figure*}[th!]
\centering
\includegraphics[width=\linewidth]{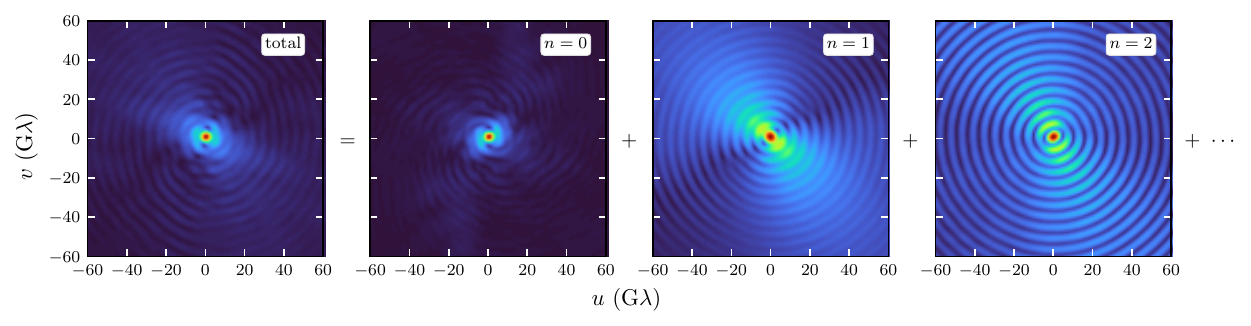}
\caption{The amplitude of the Fourier transform (visibility amplitude) of the simulated image shown in Figure~\ref{fig:nring_splash_cartoon} for the total image and its subimage decomposition. The ringing features become more prominent for higher-order subimages, reflecting the increasingly sharp and regular ring structure seen in the images.
}
\label{fig:FT_maps}
\end{figure*}

Given $C_0(\Delta t)$, it is possible to calculate the autocorrelation of the total light curve
\begin{align}
\mathcal{A}(\Delta t) =& \left\langle L_{\rm tot}(t) L_{\rm tot}(t + \Delta t) \right\rangle \nonumber \\
=& \ C_0(\Delta t) \left( 1 + e^{-2\gamma} \right) \nonumber \\
& + e^{-\gamma} \left[ C_0(\Delta t + \tau) + C_0(\Delta t - \tau) \right],
\label{eq:autocorr_semianalytic}
\end{align}
where the first term gauges the contribution of astrophysical correlations to the autocorrelation and the second term quantifies the general relativistic contribution \citep{cardenasavendano_2024_noechoes}. The Lyapunov exponent for polar orbits ranges between approximately $2.4$ and $\pi$ resulting in demagnification factors of $\approx 10 - 25$ for extremal ($\bhspin = 1$) to non-spinning black holes.
By using Equation~\ref{eq:autocorr_semianalytic} and considering different forms of $C_0(\Delta t)$, it is possible to explore when astrophysical autocorrelations can overwhelm the general relativistic echoes (as was the case for the autocorrelation in the right panel of Figure~\ref{fig:lightcurve_decomposition}). 

\vspace{0.5em}

Figure~\ref{fig:semianaytic_correlations} shows how three different hypothetical models for the astrophysical correlation $C_0(\Delta t)$ could influence the detectability of the echo signature. The top panel of that figure shows three different models for $C_0(\Delta t)$ with increasingly smooth fall-off behaviors. Model A has $C_0(\Delta t) = e^{-\left|t\right|/2\tau}$; model B has $C_0(\Delta t) = 1 - \left|t\right|/2\tau$; model C has $C_0(\Delta t) = e^{-2t^2/4\tau}$. The second panel shows the autocorrelation one would measure from Equation~\ref{eq:autocorr_semianalytic}.
A prominent peak appears in $\mathcal{A}(\Delta t)$ when $C_0(\Delta t)$ falls off quickly. But if $C_0(\Delta t)$ is smooth and decays more slowly, the echo signature can be very hard to identify in the autocorrelation alone. The last panel of the figure shows the correlation one could measure given direct access to the $n=0$ and $n=1$ light curves,
\begin{align}
\mathcal{I}(\Delta t) \, &\equiv \left\langle L_0(t) L_1(t + \Delta t) \right\rangle \\ 
&= e^{-\gamma} C_0(\Delta t - \tau).
\end{align}
Evidently, this correlation is just a time-translated copy of $C_0(\Delta t)$.

For comparison, we also include an estimate of $C_0(\Delta t)$ measured from the direct $n=0$ light curve of the {\tt{athenak}} simulation, which is even less promising than the pessimistic model B. Astrophysical events like flares may lead to transient temporal behavior that is more amenable to measuring echoes directly in the autocorrelation function \citep{wong_2021_glimmer}, but the base variability present in our GRMHD simulations seems unlikely to yield clearly observable echoes.

\section{Measuring Light Echoes}
\label{sec:observations}

While it may be challenging to detect light echoes directly in the autocorrelation function of the light curve, the bottom panel of Figure~\ref{fig:semianaytic_correlations} shows that the echo signature should still be present if it is possible to independently recover the $n=0$ and $n=1$ light curves. In this section, we show how interferometric data can be used to construct a proxy for the $n=1$ light curve and argue that the correlation of the total light curve with the proxy $n=1$ light curve should present a clear echo signature. After describing the theoretical preliminaries, we show the result of performing such a correlation on data from our GRMHD simulation.

\subsection{Interferometric signatures}
\label{sec:observations_example}

Interferometers measure complex visibilities along baselines between pairs of receivers. Given a baseline $\vec{u}$, which is a dimensionless vector (measured in units of the observation wavelength) projected orthogonal to the line of sight, an interferometer will measure the complex visibility 
\begin{align}
V\left(\vec{u}\right) = \int I\left(\vec{x}\right) e^{-2\pi i \, \vec{u} \cdot \vec{x} } \, d \vec{x},
\end{align}
where $I\left(\vec{x}\right)$ is the image of the source on the sky and $\vec{x}$ is the corresponding dimensionless image coordinate (measured in radians).

\begin{figure*}[th!]
\centering
\includegraphics[width=\linewidth]{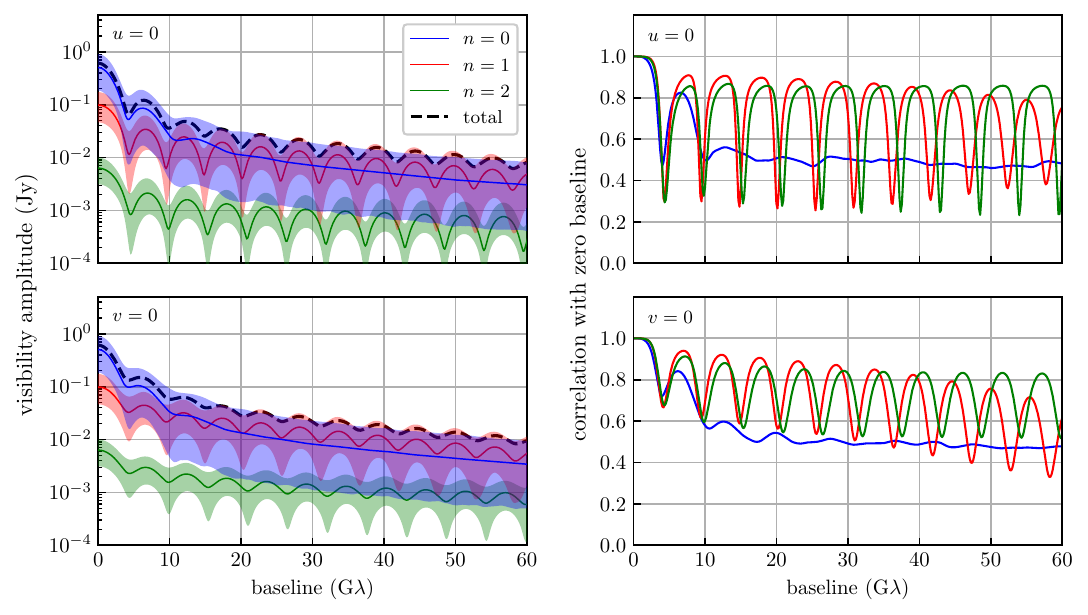}
\caption{\textit{(left)} Vertical (top) and horizontal (bottom) cross sections of the amplitude of the Fourier transform (visibility amplitude) shown in Figure~\ref{fig:FT_maps} for the total image (dashed black curve), the direct subimage ($n=0$, blue), as well as the $n=1$ (red) and $n=2$ (green) subimages. The curves show the incoherent average (i.e., the average of the visibility amplitude rather than the complex visibility) over about $20,000$ snapshots spanning about $10,000\,GM/c^3$. The shaded bands show the $\pm 2\sigma$ deviation across the time series. The variability in the $n=0$ subimage is high, which washes out the ringing structure as the locations of the nulls change over time (see e.g., \citealt{medeiros_2018_variability}). 
\textit{(right)} Vertical and horizontal cross sections of the (self-)correlations (correlation between the flux at zero baseline and the flux at longer baselines with zero time delay) for the same simulation as the left panel. The higher-order subimages show considerably higher (self-)correlation than the direct image.
}
\label{fig:visamp_slices}
\end{figure*}

The primary issue with detecting the echoes identified in Sections~\ref{sec:echo_theory}~and~\ref{sec:echo_model} is that the $n=0$ and $n=1$ signals mix together in the total image. However, under astrophysically reasonable conditions, interferometry provides a way to obtain a proxy for the $n=1$ light curve. Figure~\ref{fig:FT_maps} shows the visibility amplitudes $\left| V \left(\vec{u} \right) \right|$ in the $\vec{u} = \left( u, v\right)$ plane that an interferometer would measure for the black hole image in the right panel of Figure~\ref{fig:nring_splash_cartoon}. The figure also shows the (normalized) visibility amplitudes for the $n=0, 1, 2$ subimages. Since each higher-order subimage must be entirely contained within an increasingly narrow lensing band, the visibility amplitudes of the higher-order subimages display increasingly regular behavior \citep{johnson_2020_universal}.

The solid lines in the left panel of Figure~\ref{fig:visamp_slices} show the visibility amplitudes for the full image and subimages evaluated along the vertical ($u=0$, positive $v$) and horizontal ($v=0$, positive $u$) slices of Figure~\ref{fig:FT_maps}. The direct emission dominates the power at short baselines; however, since the $n=1$ image is thinner than the direct one, the $n=1$ signal drops off more slowly with increasing baseline compared to the $n=0$ signal. Since the contribution of the higher-order subimages to the total signal increases with increasing baseline, long baseline interferometric measurements can act as a probe of the $n=1$ signal. The difference in behaviors between the vertical and horizontal slices is due to the orientation of the underlying image---the brightness asymmetry across the photon ring differs as a function of position angle in the image (see Figure~\ref{fig:nring_splash_cartoon}).

Although the signal at longer baselines probes the higher-order subimages, the temporal variability in that signal need not be strongly related to the zero-baseline behavior, and it may seem that there are few \emph{a priori} reasons for the two time series to be related. What, then controls the variability at longer baselines? First, note that the normalization of the visibility amplitude signal is driven by the zero-baseline amplitude. Beyond the normalization, some contribution to the temporal variability will come from changes in the spatial distribution of emission across the images. However, since the higher-order images are confined to lie within their corresponding lensing bands, the magnitude of their structural variation is limited. Among the most important contributing factors to the variability are the diameter and width of the ring, which change the location of the nulls and the rate at which the signal falls off, respectively. Since the radial structure in the emissivity distribution is exponentially demagnified with each subring order, however, variations in both the ring diameter and ring width are limited. 

\begin{figure*}[th!]
\centering
\includegraphics[width=\linewidth]{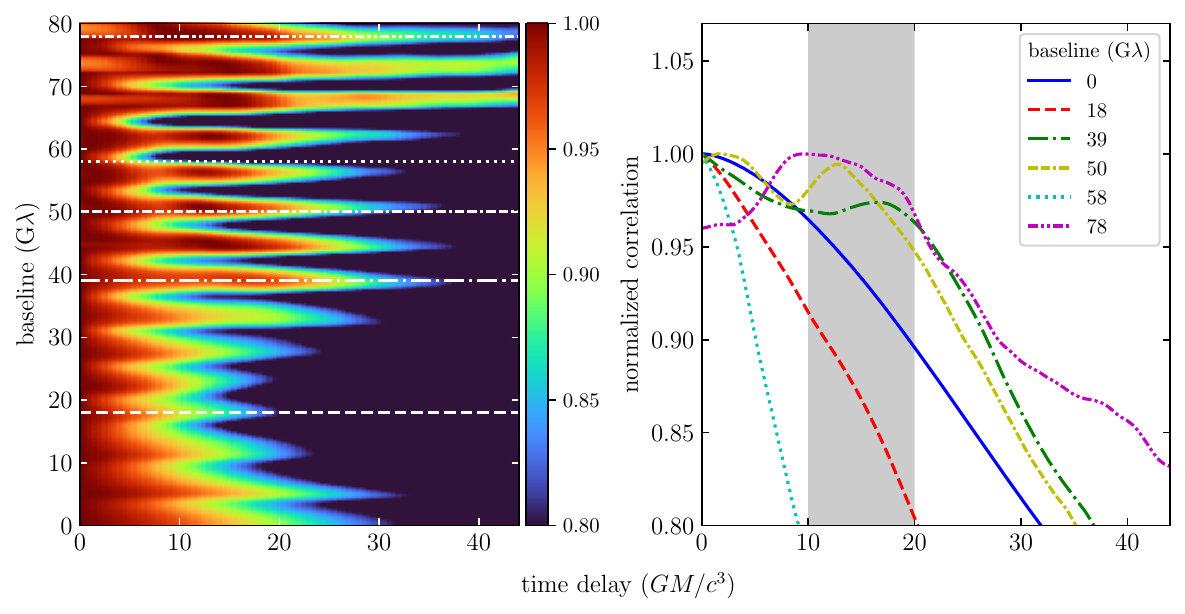}
\caption{
Demonstration of echo detection using the correlation method described in this paper applied along the $v=0$ axis of the data seen in previous figures.
\textit{(left)} The color map shows the correlation between the zero-baseline time series and the time series at a longer baseline ($y$-axis) as a function of time delay offset between the two time series ($x$-axis).  As baseline length is increased beyond about $40\,{\rm G}\lambda$, a second peak in the correlation signal appears near $\Delta t = 10-20\,GM/c^3$. \textit{(right)} We highlight several horizontal cross sections from the left panel (with matching line styles). The echo peak at 10 to 20 $GM/c^3$ is visible starting around $\sim 40\,$G$\lambda$ for this particular model (see the dark gray shaded region).
The correlations shown in this figure were performed over time series of total length $2000\,GM/c^3$.
}
\label{fig:correlations_measured}
\end{figure*}

The colored shaded regions in the left panel of Figure~\ref{fig:visamp_slices} show the envelope of variations in the visibility amplitudes over time. The envelopes for the higher order subimages show a clearer ringing structure, which is consistent with the geometric limits imposed by the lensing band.
The right panel of Figure~\ref{fig:visamp_slices} shows the correlation coefficient computed between the visibility amplitude time series at each baseline length and the zero baseline (i.e., total flux) light curve for each independent subimage. While the correlation for the direct $n=0$ image falls off quickly, the higher order $n=1$ and $n=2$ images are more strongly correlated at longer baselines, in agreement with the expectation given above motivated by geometric arguments. The correlation decreases near nulls in the underlying visibility amplitude since small variations in the ring geometry lead to large variations in the visibility amplitudes there \citep{medeiros_2018_variability}.

\vspace{0.5em}

The left panel of Figure~\ref{fig:correlations_measured} shows the temporal correlation as a function of baseline position for the data used in the previous Figures. The colors in the panel show the correlation coefficient evaluated between the visibility amplitude time series from a given point $u$ (along the $v=0$ axis) with the time series light curve of the total signal as a function of time delay lag (horizontal axis). In this format, echoes should appear as secondary bumps in the correlation at non-zero time delay. Several echoes are visible in this panel at $\Delta t \approx 14-16\,GM/c^3$ starting at baseline lengths of $\approx 40\,{\rm G}\lambda$. 

The right panel of Figure~\ref{fig:correlations_measured} shows the same (normalized) correlations as a function of time delay evaluated at different baselines locations. We have selected baselines that show how detecting the echo signature may require targeted observations, since short baselines and baselines that fall within a null will not necessarily show strong evidence of the light echo peak (e.g., there is a significant difference between the signals at $\left|u\right| = 50\,{\rm G}\lambda$ and $58\,{\rm G}\lambda$). For baselines that do exhibit the echo, there is evidence of a peak near the expected delay period $\approx 16\,GM/c^3$, although its center delay time and width change for different measurement choices. The amplitude of the correlation at the echo time delay is not always significantly larger than signal at zero time delay; thus, the difference in the signals measured along similar baselines may build confidence in an echo detection. The details of an observing strategy depend on the technical characteristics of the instrument and are beyond the scope of this paper.

\subsection{Observational considerations}

Our analysis has focused on system parameters consistent with those inferred for the black hole at the center of the M87 galaxy. The M87 accretion system is an appealing target for an echo search because it is large on the sky, varies slowly enough to enable resolved time series VLBI data, and its presumed inclination likely softens the confounding effects of non-axisymmetry and inclination on the temporal delay structure (see Appendix~\ref{app:inc_asymp}).

The method we describe requires interferometric detections along a single baseline, which is significantly easier than full imaging. It also has several other significant advantages:
\begin{itemize}
\item Since we use time series data, there is no need to record absolute phase information (as might be required when averaging to enable shadow shape measurements from the ringing behavior in the higher-order subimages).
\item The required observation cadence is not particularly restrictive: the characteristic timescale for M87 is $GM/c^3 \approx 8.9\ {\rm hr}$ ($M = 6.5\times10^9 M_\odot$) and the characteristic timescale for \sgra~is $GM/c^3 \approx 20\ {\rm sec}$ ($M = 4.14 \times 10^6 M_\odot$). The EHT has already demonstrated the technical capability to perform daily snapshot imaging on M87, and we require only an interferometric detection.
\item The total duration of the observations is reasonable: the analysis in Section~\ref{sec:observations_example} was performed over $2000\,GM/c^3$, which is two years for M87 or $11$ hours for \sgra. Additional complications arise due to \sgra's short dynamical timescale, which is significantly shorter than either the Earth's rotational period or the orbital period of a putative satellite. The approach we describe requires that approximately the same $uv$ point be sampled consistently over hundreds of dynamical times, which can be challenging for \sgra~as the Earth rotates. This lies in contrast to the case for M87, for which the timescale over which the baselines change is shorter than both the sampling timescale and the total observation duration.
\end{itemize}

Nevertheless, making an interferometric detection of echoes will require observing along longer baselines than are currently available to the EHT. The exact baselines required likely vary from model to model, and a full parameter survey to estimate likely detection rates is beyond the scope of this paper. However, based on our analysis, we infer that measurements at $\gtrsim 40\,{\rm G}\lambda$ would be required for an echo detection. Longer baseline measurements can be achieved through a combination of observing at higher frequencies and longer physical baselines, e.g., via space-VLBI. The EHT has already developed and begun installing multi-band receivers that enable $345\,$GHz observations, providing a $1.5$ times increase in accessible baselines \citep{johnson_2023_ngehtscience,raymond_2024_eht345}. Efforts to expand the EHT array into space are also well underway, with various mission concepts and pilot hardware studies being the focus of concerted scientific development \citep{kudriashov_2021_ehi,gurvits_2021_theza,kurczynski_2022_ehe,johnson_2024_bhex}. 

Obtaining a clean measurement from visibility amplitudes at long baselines due to lensed emission requires more than just access to the appropriate baseline length. For example, observations of the galactic center at long baseline are subject to diffractive scattering and refractive noise that blur and add noise to the signal, respectively. Furthermore, the effects of scattering are anisotropic with significantly greater impact on long east-west baselines compared to the north-south direction \citep{rickett_1990_scattering,narayan_1992_diffrefr,johnson_2016_refractive,johnson_2016_scattering}. Since the strength of the echo signature varies over position angle in $uv$ space (see Figure~\ref{fig:visamp_slices}), detecting echoes from \sgra~may depend on the position angle of the projected black hole spin axis on the sky and the relative orientation of the matter producing the observed emission.

Figure~\ref{fig:visamp_slices} suggests that an echo detection might require sensitivity on the order of $1 - 10\,{\rm mJy}$, which is likely feasible given the thermal noise expected for a modest baseline from ALMA to a 4-meter orbiter with 32 GHz of averaged bandwidth and a 10 minute coherent integration (see, e.g., Appendix C of \citealt{johnson_2020_universal}). Multifrequency observations may provide additional information about the echo signature, since the echoes are an achromatic prediction of general relativity. Simultaneous lower frequency observations may also enable significantly longer integration times at higher frequencies (and thus improve the noise and sensitivity limits) via frequency phase transfer methods \citep{rioja_2015_fpt,issaoun_2023_ngeht86}.

\section{Discussion}
\label{sec:discussion}

\begin{figure}[th!]
\centering
\includegraphics[width=\linewidth]{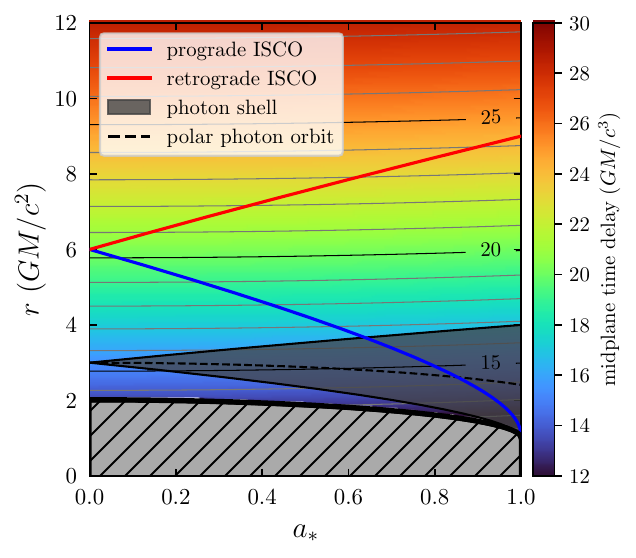}
\caption{
Echo delay seen by a face-on observer as a function of black hole spin and midplane emission location. The gray, hatched region covers the radii interior to the event horizon. The colormap shows the $n=0\to n=1$
echo delay for the emitted material as a function of black hole spin $\bhspin$ and radial position within the midplane $r$. Blue and red lines show the locations of the innermost stable circular orbit (ISCO) for prograde and retrograde massive particles, respectively. The black shaded region denotes the photon shell, extending from the prograde photon orbit (smaller radius) to the retrograde one (larger radius). The polar photon orbit, i.e., the one visible to a face-on observer, is denoted with a dashed, black line. 
The echo delay period is more sensitive to the emission location than the spin.
}
\label{fig:spin_dependence_midplane}
\end{figure}

Black hole light echoes are produced when photons emitted from the same source in different directions follow different paths to an observer. We have proposed a method to detect black hole light echoes using sparse very long baseline interferometric observations. We have used simulated data to demonstrate the viability of our method and shown that the echo signature is detectable even when the emission is produced from a mildly stochastic MAD flow. If more impulsive emission were produced, e.g., due to flaring behavior, the echo signature could be even easier to detect. For the model of M87 that we have considered, recovering the interferometric echo would require a detection on the order of a few to tens of ${\rm mJy}$ on a baseline of $\gtrsim 40\,{\rm G}\lambda$. Such a measurement should be possible with a modest space-based interferometer.

The echo time delay is roughly determined by how long it takes for light emitted near the black hole to travel around its far side, which in turn depends on the black hole mass and angular momentum as well as the inclination angle of the observer. Figure~\ref{fig:spin_dependence_midplane} shows the characteristic echo delay produced by emission at different radii in the midplane as a function of black hole spin, assuming a face-on observer. As can be seen in that figure, the echo delay period is not particularly sensitive to spin directly, although the spin may influence the location of emission. In accretion models consistent with EHT data, the majority of emission typically comes from $\approx 3\,GM/c^3$, suggesting that the characteristic echo delay period should be $\approx16\,GM/c^3$.

\begin{figure*}[th!]
\centering
\includegraphics[width=\linewidth]{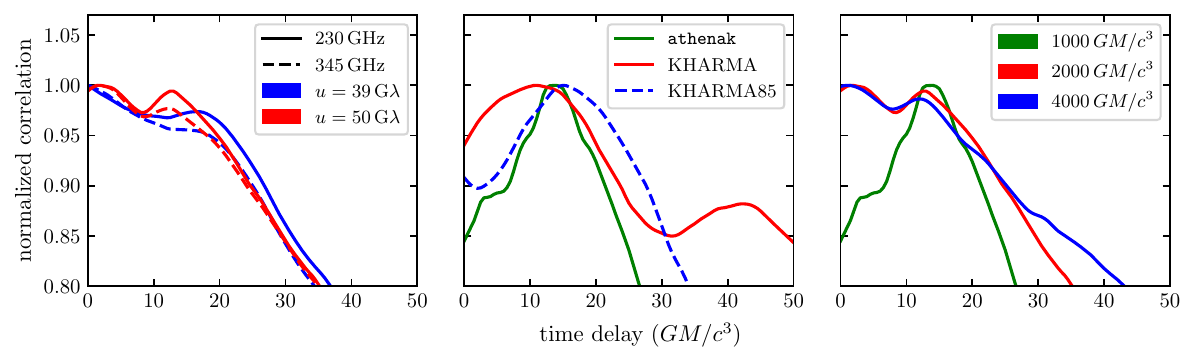}
\caption{Similar to the right panel of Figure~\ref{fig:correlations_measured} but exploring different model parameters. \textit{(left)} The normalized correlation between the total light curve and the visibility amplitude time series at different baseline lengths (red and blue) for observational frequency 230~GHz (solid curves) and 345~GHz (dashed curves). The echo signature is mildly suppressed in the $345\,$GHz observations because the long baseline signal is less correlated with the zero baseline time series (see text for more detail).
\textit{(center)} Similar to the left panel, but for different GRMHD simulations and radiative transfer parameters over the same $1000\,GM/c^3$ time window. The KHARMA85 simulation has $\bhspin = 0.85$ and $r_{\rm low} = 10$ (the remaining parameters match the other simulations). An echo signature is detectable in each simulation.
\textit{(right)} Similar to the other panels, but here we explore whether the echo signal can be detected in different windows of the full simulation. We consider non-overlapping windows with different durations (color) from the full time series. Although the echo signature differs across the windows, it is present in each of the correlation functions.
}
\label{fig:alternate_models}
\end{figure*}

For an inclined observer, emission from different parts of the accretion flow will lead to different echo delays, i.e., different isodeltachrones will be sampled by the emission, exciting an analog of black hole glimmer (see \citealt{wong_2021_glimmer} for more detail).
In principle, a clear, statistically rigorous detection of multiple echo periods could be used to constrain black hole mass, spin, and inclination as well as the statistical properties of the accretion. It could also be used to test the Kerr hypothesis, since deviations from Kerr change the expected echo signature. A full analysis of how the echo signature changes in non-Kerr metrics is beyond the scope of this paper.

The analysis presented in this paper was performed at the original $230\,{\rm GHz}$ operational frequency of the EHT; however, technological advances and upgrades to the array have enabled simultaneous measurements at $345\,{\rm GHz}$. Since echoes are an achromatic prediction of general relativity, the echo signature should be observable at any frequency, and a multi-frequency analysis could be used to build confidence in the robustness of a detection. Unfortunately, there is no direct relationship between the strength of the echo signature and observing frequency. In the left panel of Figure~\ref{fig:alternate_models}, we show the echo signature computed for the fiducial simulation for two different baseline lengths and at both $230\,{\rm GHz}$ and $345\,{\rm GHz}$. For this choice of model parameters, the correlation echo is \emph{weaker} at $345\,{\rm GHz}$, even though the higher-frequency images have decreased optical depth and the relative strength of the $n=1$ signal over $n=0$ is larger at long baselines. In this case, the echo signature is weaker due to the inherently lower correlation between the $n=1$ signals at zero and long baselines, which is because of the structure of the emission region.

While the main text focused on a single simulation, we test the robustness of our procedure by analyzing two more high-cadence black hole accretion simulations. These two simulations were run with a different GRMHD code, KHARMA \citep{prather_2021_iharm3d}, for black holes with spins $\bhspin = 0.9375$ and $\bhspin = 0.85$. Like the fiducial simulation, the KHARMA simulations were initialized for the MAD state and with a Fishbone-Moncrief torus with $r_{\rm in} = 20\,GM/c^2$, $r_{\rm max} = 41\,GM/c^2$, and $\hat{\gamma} = 13/9$. The simulations were run in a domain extending to $r = 1000\,GM/c^2$ on logarithmic spherical grids with resolutions of $384\times 192\times 192$ and $448\times 224\times 224$ in the radial, elevational, and azimuthal directions.

As with the fiducial {\tt{athenak}} simulation, the raytracing was performed using the slow-light version of {\tt{ipole}} at an inclination of $17^\circ$ and rotated to align the projected axis of the jet with large-scale observations. The $\bhspin = 0.937$ simulation used the same electron temperature parameters ($r_{\rm low} = 1$ and $r_{\rm high} = 40$) while the $\bhspin = 0.85$ simulation used $r_{\rm low} = 10$ and $r_{\rm high} = 40$. The results of performing the correlation analysis on the different simulations are shown in the center panel of Figure~\ref{fig:alternate_models}. Each of the correlation functions shows a clear echo signature at around $15-20\,GM/c^3$.

The correlation peaks seen for the KHARMA simulations appear more smooth and broad than the one for the {\tt{athenak}} simulation. This difference is driven primarily by the choice of measurement window rather than by structural differences in the underlying flows or their autocorrelation functions. The KHARMA simulations were run for a shorter time than the {\tt{athenak}} simulation. To provide a more direct comparison, we limit our analysis to observing windows with a duration of $1000\,GM/c^3$ for each simulation. Although it is possible to select a window from the {\tt{athenak}} simulation that better matches the broad peaks seen in the KHARMA simulations, we show a randomly selected window to provide a sense for the theoretical uncertainty in the structure of the correlation function.

The right panel of Figure~\ref{fig:alternate_models} shows the result of searching for echoes in the fiducial {\tt{athenak}} simulation in various subwindows of the full simulated data. The subwindows were chosen to cover different, non-overlapping observing periods, and spanned $1000$, $2000$, and $4000\,GM/c^3$. For each of the time windows we have selected, the correlation functions show clear evidence of an echo peaking between $15-20\,GM/c^3$ along a baseline between $40$ and $60\,{\rm G}\lambda$. We caution, however, that we have not exhaustively searched all possible subsets of the data and there may be some subwindows in which an echo cannot be easily detected.

\vspace{0.5em}

Our analysis is subject to several limitations including that we have assumed a nearly face-on inclination for the black hole, which limits the echo delays that can be observed. We have also only considered MAD accretion with choices for the electron thermodynamics that tend to produce emission closer to the midplane of the spacetime (see, e.g., Figure~\ref{fig:isodeltachrones_main}). With a different accretion flow, different plasma parameters, or a different viewing inclination, the echoes might be significantly harder to detect since the emission could be spread over a larger set of delays. Furthermore, if the plasma is optically thick, then the $n\ge1$ higher-order images could be more suppressed, making an echo detection more challenging.

\vspace{0.5em}

One interesting possible extension to the work presented in this paper would be to search for the echo signature in data from shorter baselines like those accessible to Earth-based observatories. We considered the visibility amplitude time series on longer baselines as a proxy for the $n=1$ signal. However, it may be possible to ``subtract off'' the $n=0$ variability from the intermediate Earth baselines using information from the total light curve. One could then search for an echo signature by correlating the residual at the intermediate baselines and the total light curve.

\vspace{0.5em}

In this paper, we have only considered signals in total intensity, but echoes should also be present in polarimetric data. \citet{palumbo_2023_ringpolarimetry} showed that the polarimetric phase should undergo a characteristic shift on long baselines; such a shift would provide compelling evidence for the presence of strongly lensed emission. It may be possible to leverage information about the polarimetric phase alongside polarimetric temporal correlations to strengthen confidence in an echo detection. Such an analysis would require a more robust understanding of Faraday effects.

Finally, we note that we have only considered correlations between the total light curve and a single, fixed point in the $uv$ plane. There are, however, informative temporal (and spectral) correlations across the entire image plane \citep{wong_2021_glimmer,hadar_2021_autocorrelations,hadar_2023_spectrotemporal}.
The spatial-temporal correlation structure of this black hole glimmer can be used to constrain system parameters, and it is likely that the \emph{interferometric} analog of glimmer produces informative correlations between different points in $uv$ space. An analysis of visibility amplitude time series across different points in the $uv$ plane (as may be accessible to a multi-spacecraft interferometer) may thus provide significant constraining power. We defer such a study to future work.

\begin{acknowledgements}
The authors thank Andrew Chael, Charles Gammie, Delilah Gates, Monika Mo\'{s}cibrodzka, and Eliot Quataert for fruitful discussions.
The authors also thank the anonymous referee for their helpful and clarifying suggestions.
G.N.W.~was supported by the Taplin Fellowship and thanks the Niels Bohr International Academy for their hospitality.
L.M.\ gratefully acknowledges support from a NASA Hubble Fellowship Program, Einstein Fellowship under award number HST-HF2-51539.001-A. 
A.C.-A. is supported by LANL Laboratory Directed Research and Development, grant 20240748PRD1, as well as by the Center for Nonlinear Studies. LANL is operated by Triad National Security, LLC, for the National Nuclear Security Administration of the U.S. DOE (Contract No. 89233218CNA000001). This work is authorized for unlimited release under
LA-UR-24-28781. Some of the simulations presented in this work were made possible by ACCESS allocation PHY220049.
\end{acknowledgements}


\appendix

\section{Inclination and the asymptotic $n$ limit}
\label{app:inc_asymp}

\begin{figure*}[th!]
\centering
\includegraphics[width=\textwidth]{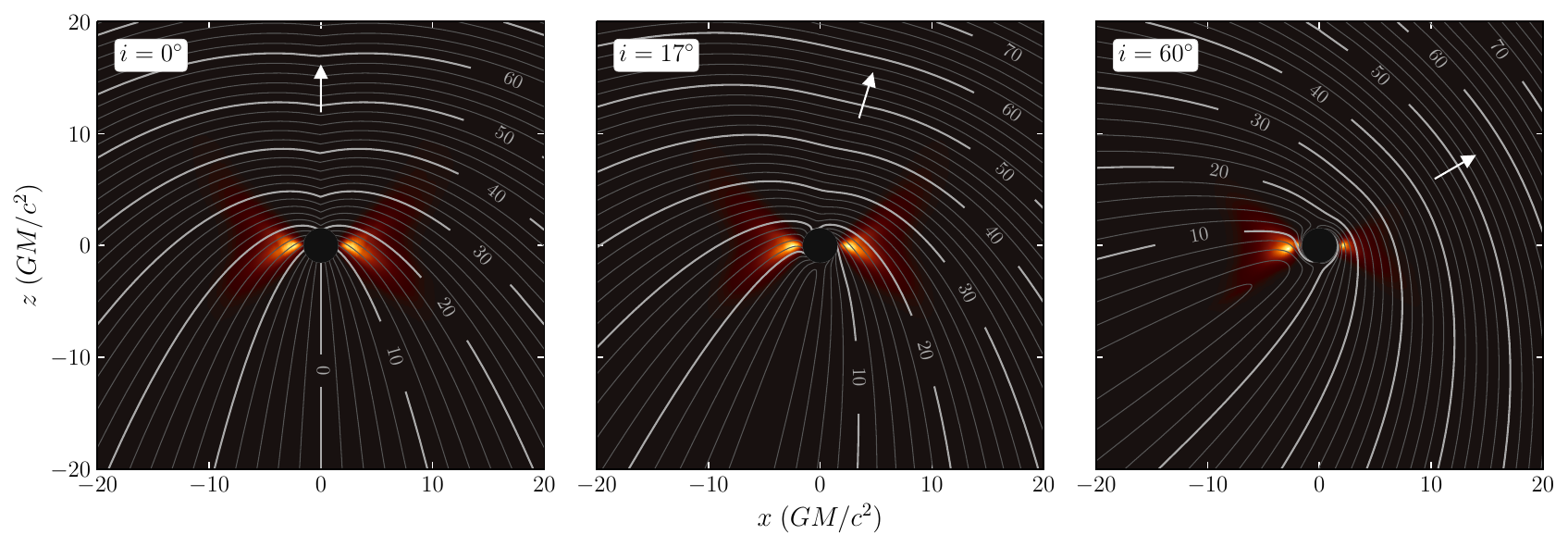}
\caption{Isodeltachrones and spatial distribution of observed emission similar to Figure~\ref{fig:isodeltachrones_main} but for different observer inclination angles $i=0^{\circ}, \, 17^{\circ}, \, \mathrm{and} \, 60^{\circ}$. The emissivity typically peaks in regions where the time delay is between $10-20\,GM/c^3$; however, for the higher inclination $i=60^{\circ}$ simulation, the emission also peaks in a region with a time delay $\sim 5 - 10$ $GM/c^3$. Unlike in Figure~\ref{fig:isodeltachrones_main}, we do not azimuthally average the emission maps so that the left-right asymmetry due to Doppler boosting is visible. 
}
\label{fig:isodeltachrones3}
\end{figure*}

\begin{figure*}[th!]
\centering
\includegraphics[width=\textwidth]{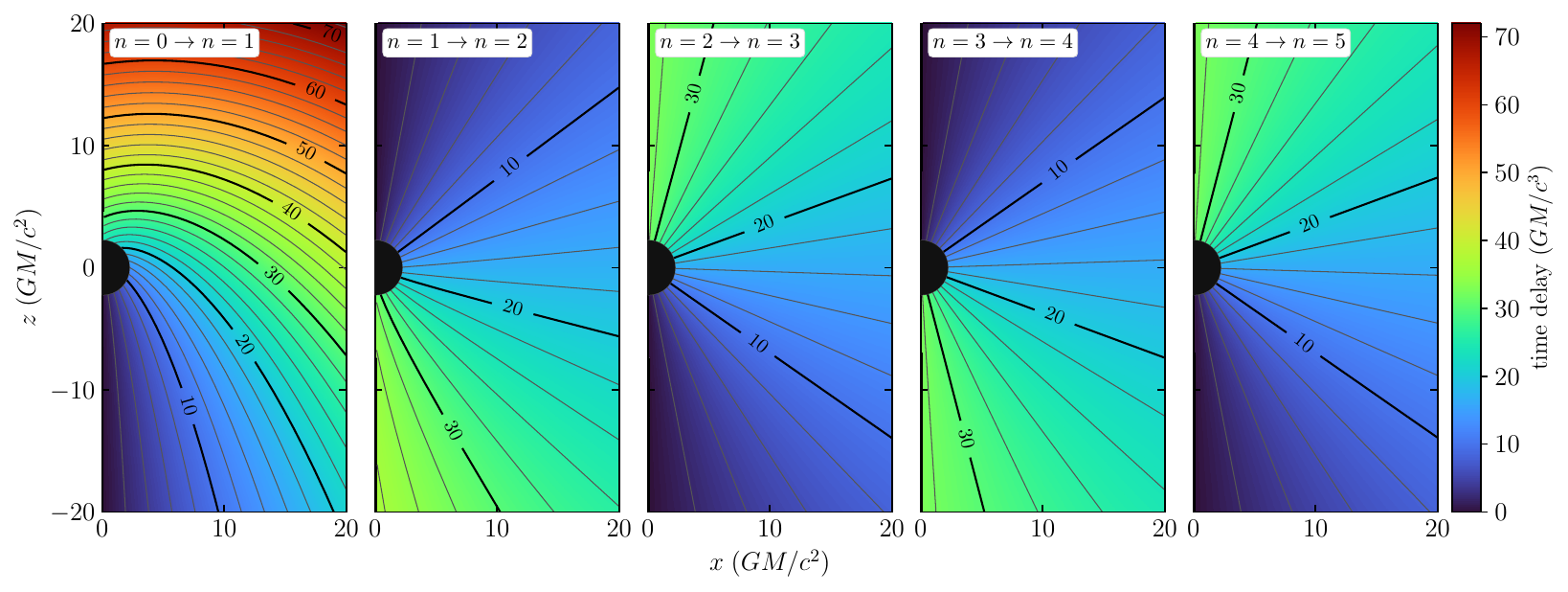}
\caption{Isodeltachrones for a Schwarzschild black hole. Both the contours and the color map show the time delay between the direct and first order indirect emission \textit{(left)}, between the first and second order indirect emission \textit{(second)}, and so on. The observer is located directly above the black hole at infinity. The zero-time-delay isodeltachrones coincide with the caustics along the $z-$axis. The isodeltachrone structure quickly converges to a universal pattern that alternatively mirrors across the $x-$axis for each higher-order pair of subimages.
}
\label{fig:isodeltachrones5}
\end{figure*}

In the main text, we focused on a fiducial model with the inclination inferred for the M87 black hole ($i=17^{\circ}$ away from the observer's line of sight; \citealt{walker_2018_m87jet}). In Figure~\ref{fig:isodeltachrones3} we show how Figure~\ref{fig:isodeltachrones_main} would change for different inclinations. The directly face-on model ($i=0^{\circ}$) is quite similar to the $i=17^{\circ}$ model, i.e., the emission peaks in regions where the time delay is between $10\,GM/c^3$ and $20\,GM/c^3$. However, for a higher-inclination model ($i=60^{\circ}$) we see significant differences: the emission peak on the far side of the black hole is brighter than the peak at the near side due to Doppler boosting. The stronger emission on the far side of the black hole peaks in a region where the time delay is between $\sim 4 -10\,GM/c^3$. The lesser peak on the near side is in a region with a time delay of between $10-20\,GM/c^3$. This difference in time delays is reasonable, since the portion of the lowest-order caustic surface (which has zero time delay) passing through the emission region is mostly located behind the black hole from the observer's point of view.\footnote{The general structure of caustics in the Kerr spacetime is complicated. Caustic surfaces for observers at non-zero inclinations typically take the form of a four-cusped astroid and wind around the black hole \citep{rauch_1994_caustics,bozza_2008_kerrcaustics}.} In general, different parts of the echo delay spectrum may be excited by different emission distributions, providing a probe of black hole mass, spin, and viewing inclination \citep{wong_2021_glimmer}. Although echoes are excited for any observer inclination, the echoes' interferometric correlation signature is less pronounced when the disk is viewed edge on, since emission coming from different azimuth angles around the disk will produce a continuum of time delays and smear the overall echo signature. We defer a detailed study of the effects of inclination to future work.

In Figure~\ref{fig:isodeltachrones5} we show the isodeltachrone maps for higher-order subimages for a Schwarzschild black hole ($\bhspin = 0$), i.e., the surfaces where the $n_t = k$ emission would arrive at the observer with a fixed time delay relative to the $n_t=k-1$ emission. The $n=0\to n=1$ isodeltachrones are quite different compared to those of higher order. Each consecutive order also exhibits a mirroring across the $x$-axis (e.g., in $n=1\rightarrow n=2$ the region with the lowest time-delay is on the near-side of the black hole). The isodeltachrone maps converge quickly to a universal pattern, since the photon paths must all increase by half an orbit for each subsequent order. Based on the regions of highest emissivity seen in the left panel of Figure~\ref{fig:isodeltachrones3}, the peak time delay for the higher-order subimages should be broadly comparable to the time delay between the direct and indirect images (approximately between $10\,GM/c^3$ and $20\,GM/c^3$).

\section{Subimage definitions}
\label{app:subimages}

\begin{figure*}[th!]
\centering
\includegraphics[width=0.95\textwidth]{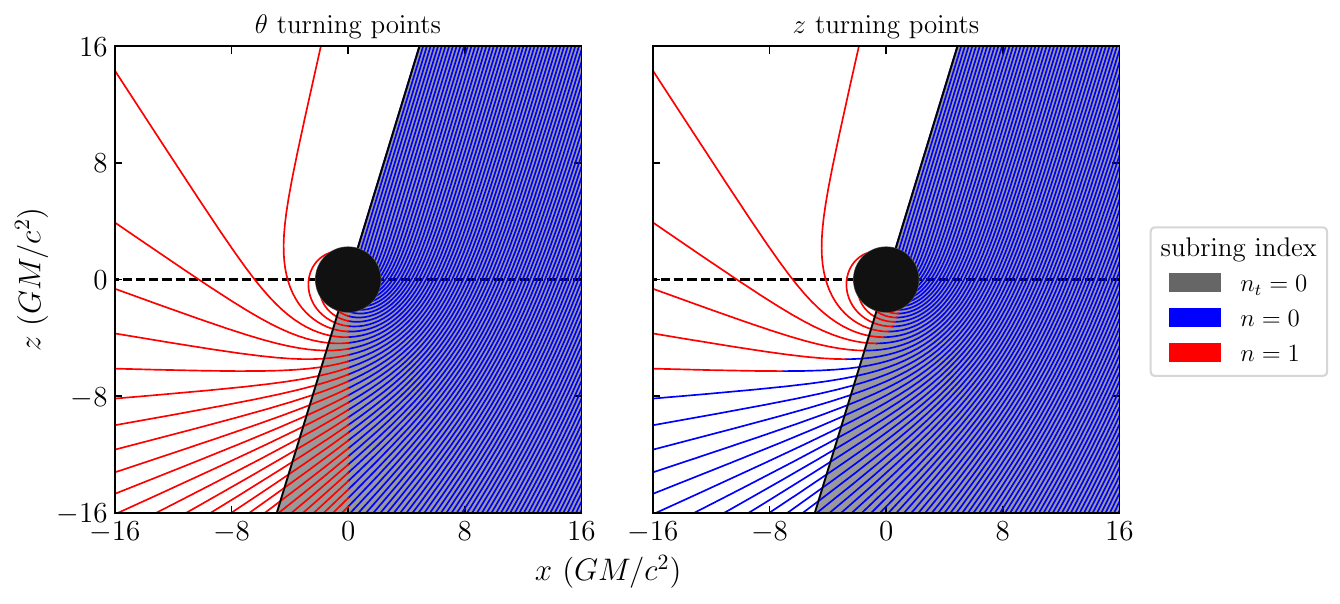}
\caption{Comparison between different subimage definitions for a non-rotating black hole. The observer is positioned at $i=17^\circ$ away from the normal to the midplane, which is shown as a black, dashed line. The colored lines represent the $\alpha = 0, \beta < 0$ geodesics that reach the observer (see Section~\ref{sec:echo_theory} for more detail). Photons emitted at different points along the same geodesic can contribute to different subimages. We color each point along the geodesics according to whether a photon emitted at that point will contribute to the direct ($n=0$, blue) or indirect ($n=1$, red) subimage. The colors in the left and right panels are evaluated according to the $d\theta/ds$ and $dz/ds$ turning-point definitions, respectively. In both panels, photons emitted along the geodesics in the underlying gray shaded region would be considered direct emission ($n_t = 0$) under the time-ordering subimage definition.
}
\label{fig:n_def}
\end{figure*}

\begin{figure*}[th!]
\centering
\includegraphics[width=0.95\textwidth]{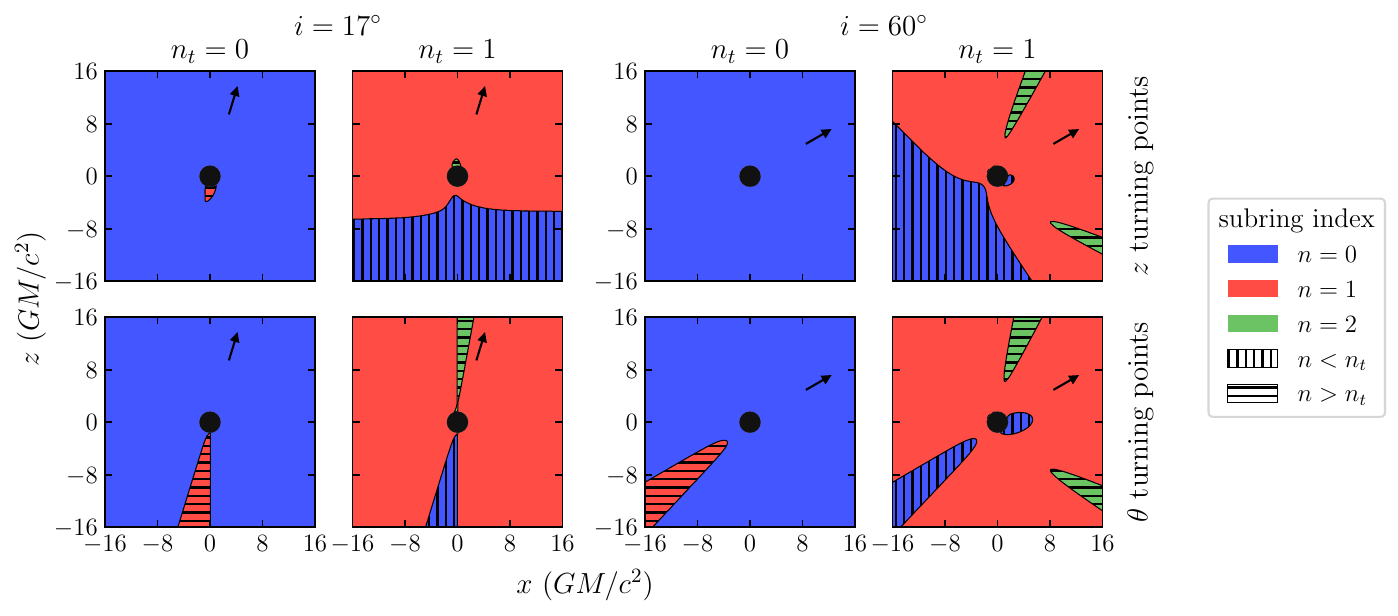}
\caption{
Comparison between subimage definitions for a black hole with spin $\bhspin = 0.9375$. Each panel considers the set of $n_t=0$ or $n_t=1$ geodesics and colors the $y=0$ plane according to their $n_z$ (top) or $n_\theta$ (bottom) classification. \textit{(top left)} At each point, we select the shortest geodesic connecting that point and an observer at inclination $i=17^\circ$. The selected geodesics at each point thus correspond to the $n_t=0$ definition. Regions where the same geodesics would be labeled $n_z=0$ are colored blue. Regions where the geodesics would be considered $n_z=1$ are red. In all panels, hashing indicates that the $n_t$ classification disagrees with the $n_z$ or $n_\theta$ classification. \textit{(top second)} Similar to the top-left panel, but here we select the second-shortest geodesics, i.e., the ones corresponding to the $n_t = 1$ definition. Regions where the same geodesics would be considered $n_z=0, 1, 2$ are shown in blue, red, and green, respectively. Points in regions that are the same color in the $n_t=0$ and $n_t=1$ panels have multiple geodesics with the same $n_z$ classification that connect them to the observer. Emission produced in these regions therefore results in multiple features within the same $n_z$ subimage.
\textit{(bottom left, bottom second)} Same as the two top-left panels but for the $n_\theta$ definition. \textit{(Four right panels)} Same as the left four panels but for inclination $i = 60^{\circ}$.
}
\label{fig:nring_definition_differences}
\end{figure*}

Strong gravitational lensing near a black hole enables light to circle around the event horizon, which allows multiple geodesics to connect every point in spacetime to an observer at a fixed position. Each geodesic can carry photons from an emission source to the observer, so a single point of emission may appear multiple times on the observer's screen. This multiple-imaging behavior suggests a conceptual decomposition of black hole images into a sequence of subimages, each comprising only photons that arrived on particular geodesics. In this appendix, we define and contrast different definitions for black hole subimages. 

\begin{figure}[th!]
\centering
\includegraphics[width=\linewidth]{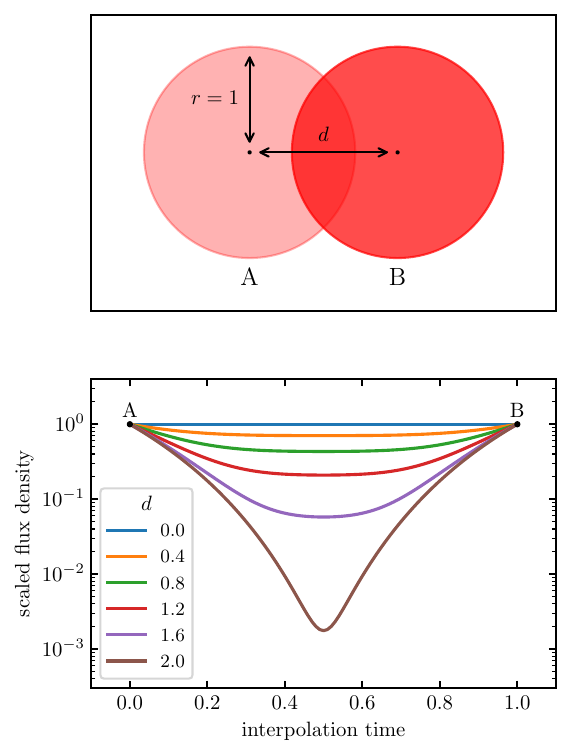}
\caption{Model of interpolation artifacts due to temporal sampling cadence. \textit{(top)} Diagram showing location of emitting sphere in snapshots A and B; the diagram shows the fluid state reconstructed by linear interpolation at $t=0.7$. The distance between points A and B scales with the velocity of the sphere. \textit{(bottom)} Demonstration of how the light curve changes with the sphere velocity.
The dip in the scaled flux density becomes more pronounced with increased sphere velocity, resulting in larger errors.
The remaining parameters for this sphere model are:
observing frequency $\nu = 230\ {\rm GHz}$,
magnetic field strength $B = 10\ {\rm G}$, 
dimensionless electron temperature $\Theta_e = 10$, and 
pitch angle between the magnetic field and photon wave vector $\theta = \pi / 3$. We assume the plasma density is low enough that the sphere is optically thin.
}
\label{fig:semianalytic_sphere_model}
\end{figure}

Unlike the time-ordering subimage definition described in Section~\ref{sec:echo_theory}, other subimage definitions in the literature associate photons with a particular subimage order by counting the number of $d\theta/ds$ ($n_\theta$) or $dz/ds$ ($n_z$) sign flips between the point of emission and the observer. When the $n_z$ definition is used, the $z$ axis is typically aligned with the black hole spin axis or, for Schwarzschild, the axis perpendicular to the accretion disk. The $n_z$ definition has the nice property that a geodesic that passes through the midplane $n$ times will encounter $n-1$ turning points in $d z / ds$.

Turning-point definitions are ``local'' to each geodesic in the sense that the subimage order for a photon emitted along a geodesic can be computed directly from that single geodesic trajectory. In contrast, the method for assigning photons to subimages described in Section~\ref{sec:echo_theory} is non-local, since it requires finding all geodesics that pass between the emission point and the observer. The time-ordered geodesic definition has the desirable property that, for a given subimage $n$, it introduces a unique mapping between each point around the black hole and the geodesics that connect that point to the observer.\footnote{By definition, this property cannot hold along caustic surfaces.} This definition is particularly conducive to an echo analysis, because it means that any source of emission can be seen at most once per subimage. This unique-mapping property is not shared by the turning-point definitions (see, e.g., \citealt{zhou_2024_forwardkerr} for a related discussion of geodesic labeling and multiple images).

The relative smoothness of the Kerr spacetime means that subimages are restricted to lie completely within geometric regions on the image plane, called lensing bands. Given an observer position and set of black hole parameters, exactly one lensing band exists per subimage order. Lensing bands may overlap because subimages are defined by a set of photons rather than their geodesic trajectories. Figure~\ref{fig:n_def} shows how photons emitted at different points along the same geodesic may contribute to different-order subimages, depending on the selected subimage order definition. In each panel, photons emitted within the blue region would appear in the $n=0$ subimage, and photons emitted in the red region would appear in the $n=1$ subimage \citep[see, e.g.,][]{cardenasavendano_2024_lensingbands}. In both panels, photons emitted along the geodesics within the shaded gray $n_t=0$ region would contribute to the time-ordered $n=0$ subimage.

Figure~\ref{fig:nring_definition_differences} shows the mapping between points on the $y=0$ plane and the subimage they contribute to for a black hole with spin $\bhspin=0.9375$. The different rows show how the mapping changes with definition, and the different pairs of columns show different viewing inclinations. The procedure for coloring each panel is as follows: 
\begin{enumerate}
\item For each point in the plane, find all geodesics connecting that point to the observer,
\item Identify which geodesic corresponds to $n_t=0$ (left) or $n_t=1$ (right), in the time-ordering definition,
\item Identify which subimage a photon emitted along that geodesic would appear in according to the $dz/ds$ (top) or $d\theta/ds$ (bottom) turning point definition.
\end{enumerate} 
Hatched regions indicate where the time-ordered and turning-point definitions would differ. In each pair of neighboring panels (i.e., for a particular inclination and subimage definition), points that lie in regions that are the same color in both panels \emph{do not} have a unique mapping between subimage order and geodesic. For example, photons emitted from $(x, z) = (8, -8)$ would appear in two places in the $n_z=0$ subimage at $i=17^\circ$ (top-left pair). Similarly, photons emitted at $(x, z) = (3, 0)$ would appear in two locations in the $n_\theta = 0$ subimage for the observer at inclination $i=60^\circ$ (bottom-right pair). Thus, when using the turning point (or midplane crossings) definitions, one should be aware of the fact that a single source may appear in multiple locations within a single subimage (see \citealt{zhou_2024_forwardkerr} for more discussion of image level degeneracy).

\section{Effects of GRMHD snapshot cadence}
\label{app:grmhd_cadence}

\begin{figure*}[th!]
\centering
\includegraphics[width=0.9\textwidth]{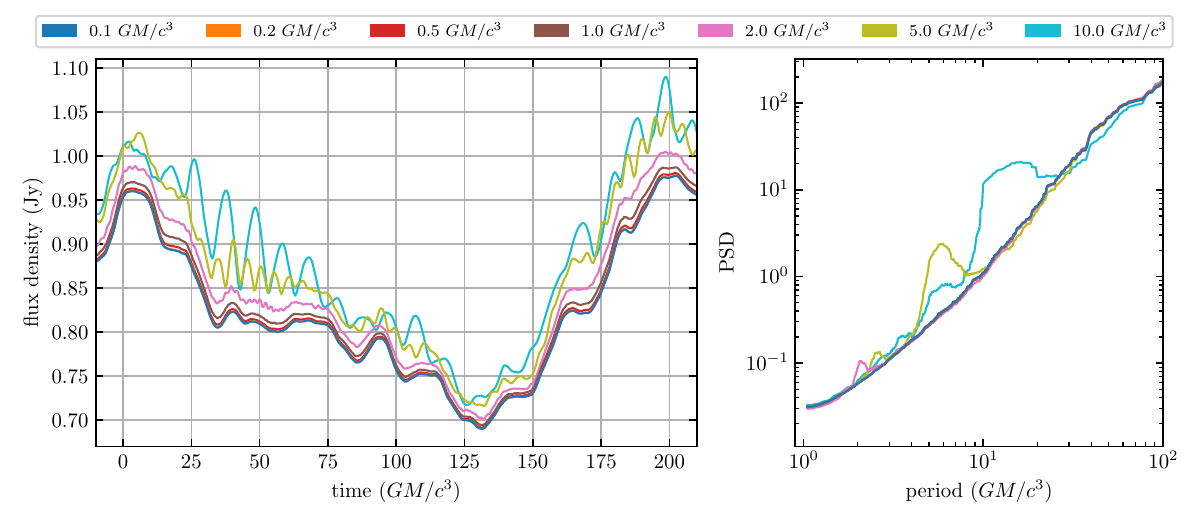}
\caption{
\textit{(left)} Light curves computed for the same $\bhspin=0.85$ GRMHD simulation but sampled with different temporal resolutions (colors). At low temporal resolutions, linear interpolation artifacts due to non-linearities in the absorption coefficients generate spurious variability and ringing features. At temporal resolutions $\lesssim 1\,GM/c^3$, the effect of the artifacts is subdominant and the light curve converges. \textit{(right)} The power spectral density (PSD) of the light curves on the left. As expected, the low-resolution light curves see considerable excess power (i.e., excess variability) due to the interpolation artifacts and converge at temporal resolutions $\lesssim 1GM/c^3$. 
}
\label{fig:grmhd_lightcurve_cadence_examples}
\end{figure*}

When computing images from GRMHD data, we use a ``slow-light'' method, which accounts for the finite speed of light and allows the fluid to evolve as photons propagate through the simulation domain. We perform the radiative transfer calculations after running the fluid simulation in a post-processing step. We use a discrete time series of GRMHD snapshots and interpolate to recover the fluid state for times between the snapshots. There is freedom in how to perform the interpolation---for example, one could envisage using the instantaneous fluid velocity to reconstruct the fluid state by propagating fluid elements throughout the domain---but a simpler implementation is to linearly interpolate in the three spatial dimensions and time independently.

Unfortunately, since the radiative transfer coefficients are non-linear in the fluid variables, linear interpolation in time introduces bumpy artifacts in the light curve as ``emission volumes'' jump between locations. This effect can be seen with a simple model: consider an optically thin, uniform ball of plasma with unit radius moving through the domain at constant velocity. The correct light curve from such an emission source should be constant in time. However, if the time series is sparsely sampled, then linearly interpolating the fluid state between snapshots A and B will lead to the slow ``fade out'' of the plasma ball in position A and the slow ``fade in'' of the ball at position B. Figure~\ref{fig:semianalytic_sphere_model} shows a schematic of the model geometry (top panel) and the calculated light curves (bottom panel) for different velocities for the plasma ball. The error in the light curve grows as the velocity of the sphere increases.

Figure~\ref{fig:grmhd_lightcurve_cadence_examples} shows light curves computed using {\tt{ipole}} in slow-light mode for different GRMHD snapshot cadences, $\Delta t = 0.1, 0.2, 0.5, 1, 2, 5,$ and $10\,GM/c^3$. The fluid data are from the same $\bhspin=0.85$ KHARMA simulation described in Section~\ref{sec:discussion}. The left panel shows the light curves, which have been computed at data points every $0.5\,GM/c^3$. Here, the unconverged light curves exhibit bumps (rather than dips as in Figure~\ref{fig:semianalytic_sphere_model}) due to optical depth effects. Since absorptivity decreases non-linearly in a manner similar to emissivity, the relative importance of absorption vs.~emission is model dependent.

The right panel of Figure~\ref{fig:grmhd_lightcurve_cadence_examples} shows the (smoothed) power spectral density (PSD) for each of the light curves in the left panel. Evidently, the light curves produced from GRMHD snapshots with longer cadence show clear artifacts at timescales commensurate with the GRMHD sampling cadence. When looking for echoes, it is particularly important that interpolation artifacts not introduce spurious features on the echo timescale. For this model, the PSD seems to converge at a sampling cadence of $1\,GM/c^3$; however, this timescale may not be representative of other models, since it depends on the structure and velocity of the emitting material. The GRMHD simulations used in this paper have snapshot cadences of $\Delta t = 0.1$ or $0.5\,GM/c^3$.

\bibliography{main}
\bibliographystyle{aasjournal}

\end{document}